\documentclass[a4paper]{article}

\usepackage{INTERSPEECH2021}

\usepackage{amsmath,graphicx}
\usepackage{cite}
\usepackage{siunitx}
\usepackage{url}
\usepackage{hyperref}
\usepackage{dcolumn}
\usepackage{pifont}
\usepackage{comment}
\usepackage{mathtools}
\usepackage{multirow}
\usepackage{tabularx}
\usepackage{subfigmat}
\usepackage[linesnumbered,ruled,vlined,commentsnumbered]{algorithm2e}

\SetCommentSty{mycommfont}

\newcommand{\trans}[1]{#1^\mathsf{T}}

\DeclareMathOperator*{\argmax}{arg\,max}
\def\appendixautorefname~#1\null{~#1 \null}
\DeclareMathSymbol{\mhyph}{\mathalpha}{operators}{`-}

\newcommand{\cmark}{\ding{51}}

\def\appendixautorefname~#1\null{~#1 \null}

\def\equationautorefname~#1\null{Eq.~(#1)\null}

\title{Online Streaming End-to-End Neural Diarization\\Handling Overlapping Speech and Flexible Numbers of Speakers}
\name{\begin{tabular}{c}Yawen Xue$^1$, Shota Horiguchi$^1$, Yusuke Fujita$^1$, Yuki Takashima$^1$, Shinji Watanabe$^2$\\ Paola Garc\' ia$^3$, Kenji Nagamatsu$^1$\end{tabular}}
\address{
$^1$ Hitachi, Ltd. Research \& Development Group, Japan\\
$^2$ Language Technologies Institute, Carnegie Mellon University, USA\\
$^3$ Center for Language and Speech Processing, Johns Hopkins University, USA}
\email{yawen.xue.wn@hitachi.com}

\begin{document}
% 最終奥義・スペース設定コマンド
% 数式の上下を詰める
\abovedisplayskip=5pt
\belowdisplayskip=5pt

% 図と図の間の余白
\setlength\floatsep{8pt}
% ページの上下に出力される図と本文の間の余白
\setlength\textfloatsep{8pt}
% ページの途中に出力される図と本文の余白
%\setlength\intextsep{0pt}
% 図とキャプションの間の余白
\setlength\abovecaptionskip{5pt}
% 2カラムにわたる図と図の間の余白
\setlength\dblfloatsep{5pt}
% 2カラムにわたる図と本文の間の余白
\setlength\dbltextfloatsep{11pt}

% 表の各列の上下の余白
%\aboverulesep=0.25ex %default: 0.4ex
%\belowrulesep=0.5ex %default: 0.65ex
\maketitle
\begin{abstract}
%This paper extends our offline end-to-end neural diarization (EEND) method, which can handle overlapping speech and flexible numbers of speakers, to online speaker diarization without re-training existing offline models.
%The original EEND model has already achieved significant improvement, especially on overlapping speech but it has no condition to decide the order of output speakers across chunks in online processing.  
%This paper solves this problem by proposing a extension of speaker-tracing buffer (STB) method that is suitable for flexible numbers of speakers named as FLEX-STB.
%The original STB utilizes the previous chunk information to trace the consistent speaker order across chunks during inference but it is limited to fixed two-speaker.
%To deal with the varying numbers of speakers across chunks, FLEX-STB applys zero-padding mechanism to increase the number of speaker slots.
%Four selection strategies are also proposed to extend the application scope of the proposed system. 
%Experiments on two real datasets: CALLHOME and DIHARD II datasets show that the proposed $\mathit{online}$ method can achieve comparable performance to the offline EEND method with \SI{1}{\second} chunk size. 
%In a \SI{10}{\second} chunk size condition, our proposed method outperforms the recently proposed block-wise encoder-decoder based EEND approach.
We propose a streaming diarization method based on an end-to-end neural diarization (EEND) model, which handles flexible numbers of speakers and overlapping speech.
In our previous study, the speaker-tracing buffer (STB) mechanism was proposed to achieve a chunk-wise streaming diarization using a pre-trained EEND model. STB traces the speaker information in previous chunks to map the speakers in a new chunk. However, it only worked with two-speaker recordings.
In this paper, we propose an extended STB for flexible numbers of speakers, FLEX-STB. The proposed method uses a zero-padding followed by speaker-tracing, which alleviates the difference in the number of speakers between a buffer and a current chunk. We also examine buffer update strategies to select important frames for tracing multiple speakers.
Experiments on CALLHOME and DIHARD II datasets show that the proposed method achieves comparable performance to the offline EEND method with 1-second latency. 
The results also show that our proposed method outperforms recently proposed chunk-wise diarization methods based on EEND (BW-EDA-EEND).
\end{abstract}

\section{Introduction}
\label{sec:intro}
Speaker diarization, a challenging technique that responds to the question ``who spoke when'' \cite{tranter2006overview,  Anguera2012, ryant2019second, ryant2020third, park2021review, horiguchi2021hitachi}, assigns speaker labels to audio regions.
Diarization produces outcomes that downstream tasks can utilize. 
For example, it can provide the turn-taking information and build a pre-processing pipeline for automatic speech recognition
in meetings \cite{kang2020multimodal,yoshioka2019advances, carletta2005ami, janin2003icsi}, call-center telephone conversations \cite{callhome,martin2000nist,senoussaoui2013study}, and home environments \cite{barker2018fifth, kanda2018hitachi, watanabe2020chime}.

%Speaker diarization should satisfy three requirements: overlapping speech, an unknown number of speakers, and online operation. 
The three challenging aspects that current speaker diarization systems should fulfill are overlapping speech, unknown number of speakers, and online operation.
However, it is still an open problem to solve these conditions at once. 
Conventional clustering-based systems primarily focus on clustering algorithms and speaker embeddings such as Gaussian mixture models (GMM) \cite{geiger2010gmm, markov2008improved}, i-vector \cite{madikeri2015integrating, garcia2017speaker,zhu2016online}, d-vector \cite{zhang2019fully, fini2020supervised}, and x-vector \cite{diez2019bayesian, mccree2019speaker}.
However, most clustering-based systems assume that there is only one speaker per segment. 
As a result, these systems cannot deal with the overlapping speech in general except for a few studies, e.g., \cite{huang2020speaker}. 

To solve the overlapping issue, an end-to-end neural diarization model (EEND) was proposed \cite{fujita2019end1}.
EEND directly minimizes the diarization error by mapping the multi-speaker mixture recording to joint speech activities using a single neural network.
The model estimates the speech activity using a dedicated stream for every speaker; hence, EEND inherently assigns two or more labels to the overlapping regions. 
EEND has already shown significant performance improvement on overlapping speech, especially after adopting the self-attention mechanism (SA-EEND) \cite{fujita2020end}, and with a fixed number of speakers.
\begin{table}[t]
    \centering
    \caption{Comparison of speaker diarization methods.}
    \resizebox{\linewidth}{!}{%
    \begin{tabular}{@{}lccc@{}}
        \toprule
        Method & Online & Overlapping & Flexible \#speakers \\
        \midrule
        x-vector+clustering \cite{diez2019bayesian}& --  & -- & \cmark\\
        UIS-RNN \cite{zhang2019fully, fini2020supervised} & \cmark  & -- & \cmark\\
        EEND/SA-EEND \cite{fujita2019end1, fujita2019end2, fujita2020end} & --  & \cmark & --  \\
        EEND-EDA/SC-EEND \cite{horiguchi2020end, fujita2020neural} & -- & \cmark & \cmark \\
        RSAN \cite{von2019all, kinoshita2020tackling} &\cmark&\cmark&\cmark\\
        BW-EDA-EEND \cite{han2020bw} & \cmark & \cmark & \cmark \\
        This work & \cmark & \cmark & \cmark\\ 
        \bottomrule
    \end{tabular}%
    }
    \label{tab:sys}
\end{table}
To deal with overlapping speech and flexible numbers of speakers, Horiguchi {\it et al.} introduced the encoder-decoder based attractor (EDA) module to SA-EEND \cite{horiguchi2020end}, and Fujita {\it et al.} extended the SA-EEND to speaker-wise conditional EEND (SC-EEND) \cite{fujita2020neural,takashima2021end}.
Both extensions have only been evaluated in offline mode.

To cope with online applications, the speaker-tracing buffer (STB)  \cite{xue2021online} was proposed to trace the speaker permutation information across chunks which enables the offline pre-trained SA-EEND model to work in an online manner.
The original STB achieved comparable diarization accuracy to the offline EEND with \SI{1}{\second} chunk size but this method was limited to two-speaker recordings.
In \cite{han2020bw}, Han {\it et al.} proposed the block-wise-EDA-EEND (BW-EDA-EEND) which makes the EDA-EEND work in an online fashion.
Motivated by Transformer-XL \cite{dai2019transformer}, this approach utilizes the previous hidden states of the transformer encoder as input to the EDA-EEND.
%However, BW-EDA-EEND only conducted experiments on \SI{10}{\second} chunk size conditions which are too large to be acceptable for the online requirement. 

To satisfy all the three requirements together, among the existing diarization methods as shown in \autoref{tab:sys}, the Recurrent Selective Attention Network (RSAN) \cite{von2019all,kinoshita2020tackling} and the block-wise-EDA-EEND (BW-EDA-EEND) stand out.
However, due to the speech separation-based training objective, RSAN is hard to adapt to real recordings, and the evaluations under real scenarios are not reported. 
On the other hand, although BW-EDA-EEND \cite{han2020bw} conducted online experiments on \SI{10}{\second} chunk size conditions, which cause large latency. In this paper, we  consider more realistic {\it streaming} applications with a smaller chunk size such as \SI{1}{\second}. 

In this work, we extend the inference algorithm of existing offline model (e.g., EEND-EDA) to operate in an online mode using the speaker-tracing buffer for flexible numbers of speakers (FLEX-STB) without re-training the offline model.
FLEX-STB is designed to deal with variable numbers of speakers using a zero-padding mechanism with reasonable latency.
Four frame selection strategies are also proposed to contain the speaker permutation information in FLEX-STB.
The proposed diarization system can operate in an \textit{online mode} handling \textit{overlapping speech} and \textit{flexible number of speakers}, and working in real scenarios such as CALLHOME and DIHARD II with \SI{1}{\second} chunk size. 

\section{Preliminary}
\label{sec:pre}
In this section, we briefly explain two key elements: EEND for flexible numbers of speakers and the original STB that enables the offline SA-EEND systems to work online. 

\subsection{EEND for flexible numbers of speakers}\label{sec:pre_eend}
Given a $T$-length sequence of $D$-dimensional log-scaled Mel-filterbank-based acoustic features $\mathbf{X}\in\mathbb{R}^{D\times T}$, a neural network-based function $\mathrm{EEND} : \mathbb{R}^{D\times T} \rightarrow(0,1)^{S \times T}$ calculates posterior probabilities of speech activities at each time frame $\hat{\mathbf{Y}}=(\hat{\mathbf{y}}_t)_{t=1}^{T}\in(0,1)^{S\times T}$ as follows:
\begin{equation}
    \hat{\mathbf{Y}} = \mathrm{EEND}(\mathbf{X}),
\end{equation}
Here, $\hat{\mathbf{y}}_t\coloneqq\trans{\left[\hat{y}_{1,t},\dots,\hat{y}_{S,t}\right]}$ is the posterior of speech activities calculated for each speaker $s\in\{1,\dots,S\}$ independently, where $\trans{\left(\cdot\right)}$ denotes the matrix transpose and $S$ is the number of speakers.
Diarization results $\tilde{\mathbf{Y}}=(\tilde{y}_{s,t})_{s,t}\in\{0,1\}^{S\times T}$ are obtained by applying a threshold value $\theta$ (\textit{e.g.}, 0.5) to the posteriors $\hat{\mathbf{Y}}$.
If $\tilde{y}_{s,t}=\tilde{y}_{s',t}=1~(s\neq s')$, it means that both speakers $s$ and $s'$ are estimated to have spoken at time $t$, which is regarded as the overlapping region.
If $\forall s\in \{1, \dots, S\},~\tilde{y}_{s,t}=0$, it indicates that no speaker is estimated to have spoken at time $t$.
Note that EEND used permutation invariant training \cite{fujita2019end1} so that there is no condition to decide the order of output speakers. 

While the original EEND \cite{fujita2019end1,fujita2019end2} fixes the number of speakers $S$ by its network structure, variants of EEND \cite{horiguchi2020end, fujita2020neural, takashima2021end} have been proposed to estimate the number of speakers $\hat{S}$. 
However, these methods perform only in the offline setting.
%This paper propose a method that can be applied to such extended EENDs to make them work online.
%To our knowledge, two extensions of EEND can perform speaker diarization of unknown numbers of speakers: EEND-EDA \cite{horiguchi2020end} and SC-EEND \cite{fujita2020neural}.
%In order to extend EEND to perform speaker diarization of unknown numbers of speakers, EEND-EDA \cite{horiguchi2020end} was proposed. 
%EEND-EDA incorporates the encoder-decoder based attractor (EDA) to the frame-wise speech embedding sequences derived from stacked Transformer encoders \cite{horiguchi2020end}. 
%A flexible number of attractors are generated by EDA, which are then multiplied by the speech embedding sequence to produce the same number of speaker activities.
%SC-EEND \cite{fujita2020neural} decodes speaker-wise speech activities sequentially conditioned on previously estimated speech activities.
%In this iterative process, the method handles variable number of speakers when an appropriate stop sequence condition is met.
%EEND-EDA can achieve excellent diarization results in the offline setting.
%In this study, we call these flexible-numbers-of-speaker EEND models FS-EEND.

%The original FS-EEND systems are designed for offline applications. 
%In an online setting, we assume that $\mathbf{X}$ is divided into short chunks arriving one by one.
%As there is no relationship among chunks, we cannot obtain a consistent speaker permutation of the whole recording, which degrades diarization performance much \cite{xue2020online}. 

\subsection{Speaker-tracing buffer for fixed number of speakers}
One of the straightforward online extensions of EEND is to perform diarization process for each chunk of acoustic features and concatenated diarization results across the chunk.
However, this cannot obtain a consistent speaker permutation of the whole recording.
This is because the EEND used permutation invariant training \cite{fujita2019end1} so that there is no condition to decide the order of output speakers.
We call this speaker permutation problem.
To solve the speaker permutation problem, we have proposed speaker-tracing buffer (STB) \cite{xue2021online} for the original EENDs, which assume that the number of speakers was known as prior.

Let $\mathbf{X}_i \in \mathbb{R}^{D \times \Delta}$ represents the subsequence of $\mathbf{X}$ at chunk $i\in\left\{1,\dots,I\right\}$ with a fixed chunk length $\Delta$, i.e., $\mathbf{X}=\left[\mathbf{X}_1,\dots,\mathbf{X}_i,\dots,\mathbf{X}_I\right]$.
The $\mathrm{EEND} : \mathbb{R}^{D\times T} \rightarrow(0,1)^{S \times T}$ function accepts the input features of flexible length $T$ and produces the posteriors of speech activities of the same length for each speaker.
Note that the number of speakers $S$ is fixed in this section.
%, where $\hat{S}_i$ is also estimated in the function.

\subsubsection{Initialization}
The STB possesses two matrices: acoustic features $\mathbf{X}^{(\text{buf})}_i\in\mathbb{R}^{D\times L_i}$ and the corresponding posteriors $\mathbf{Y}_i^{\text{(buf)}}\in\mathbb{R}^{S\times L_i}$ from $\mathrm{EEND}\left(\cdot\right)$, where $L_i$ is the buffer length after $i$-th update.
The matrices are initialized at the first chunk as follows:
\begin{align}
    \mathbf{X}_1^{\text{(buf)}} & = \mathbf{X}_1, \label{eq:init1}\\
    \mathbf{Y}_1^{\text{(buf)}} & =\hat{\mathbf{Y}}_1 = \mathrm{EEND}(\mathbf{X}_1).
    \label{eq:init2}
\end{align}
As we assume that the chunk size $\Delta$ is smaller than the maximum number of frames $L_\text{max}$ in the buffer, all the inputs and outputs of the first chunk can be fed into STB.

\subsubsection{Chunk-wise processing handling speaker permutation}
From the second chunk, posteriors $\hat{\mathbf{Y}}_i$ are computed using the STB.
Firstly, an input concatenated with the the buffer is fed into $\mathrm{EEND}\left(\cdot\right)$:
\begin{equation}
    \left[\mathbf{\hat{Y}}_{i-1}^\text{(buf)}, \mathbf{\hat{Y}}_i\right]  = \mathrm{EEND}\left(\left[\mathbf{X}_{i-1}^\text{(buf)}, \mathbf{X}_i\right]\right) \in (0,1)^{S \times (L_{i-1} + \Delta)}.
\label{eq:fs-eend}
\end{equation}
%where $\hat{S}_i$ is the estimated number of speakers of chunk $i$ and $L_{i-1}$ is the current length of the buffer. 
Next, the optimal speaker permutation for the current chunk is calculated as follows:
\begin{equation}
    \psi = \argmax_{\phi \in \mathrm{Perm}(S_i)} \mathrm{Corr}\left(\mathbf{Y}_{i-1}^{\text{(buf)}}, \mathbf{P}_\phi \mathbf{\hat{Y}}^{\text{(buf)}}_{i-1}\right), \label{eq:corr}
\end{equation}
where $\mathbf{P}_\phi \in [0,1]^{S \times S}$ is a permutation matrix for the $\phi$-th permutation in $\mathrm{Perm}(S_i)$, which is all the possible permutations of the sequence $\left(1,\dots,S\right)$.
$\mathrm{Corr}\left(\mathbf{A},\mathbf{B}\right)$ calculates the correlation between two matrices $\mathbf{A}=\left(a_{ij}\right)_{jk}$ and $\mathbf{B}=\left(b_{jk}\right)_{ij}$ defined as 
\begin{align}
    \mathrm{Corr}\left(\mathbf{A},\mathbf{B}\right)\coloneqq\sum_{i,j}\left(a_{jk}-\bar{a}\right)\left(b_{jk}-\bar{b}\right),\label{eq:correlation}
\end{align}
where $\bar{a}$ and $\bar{b}$ are the mean values of $\mathbf{A}$'s and $\mathbf{B}$'s elements, respectively.
Finally, the posterior probabilities of the $i$-th chunk are calculated with the permutation matrix that gives the highest correlation as follows:
\begin{equation}
    \mathbf{Y}_i = \mathbf{P}_\psi \mathbf{\hat{Y}}_i.
\end{equation}

If the length of $\left[\mathbf{Y}^\text{(buf)}_{i-1}, \mathbf{Y}_i\right]$ is larger than the predetermined maximum buffer length $L_\text{max}$, we select frames to be kept in the STB, which are used to solve the speaker permutation problem occurred by the future inputs.
In the paper \cite{xue2021online}, four selection strategies have been proposed.

The STB is a solution to the online diarization problem; however, it cannot be directly applied to EEND for unknown and flexible numbers of speakers.
One reason is because the number of speakers may be different across chunks so that we cannot calculate correlation using \autoref{eq:correlation}.
The other reason is that the most promising selection strategy used the absolute difference of probabilities of two speakers' speech activities; thus, the method is limited to two-speaker EENDs.

\section{Proposed method}
\label{sec:pro}
In this paper, we proposed the FLEX-STB which extends the STB coping with the two obstacles to use it with EEND for unknown numbers of speakers \cite{horiguchi2020end,fujita2020neural}.
The FLEX-STB deals with the varying number of speakers across chunks by increasing the number of speaker slots in the speaker-tracing buffer with the zero-padding in \autoref{sec:re-stb}.
When the system detects new speakers, it adds new zero-speaker-activity slots to the speaker buffer. 
We also propose four selection strategies to update the buffer, each of which are not limited by the number of speakers, in \autoref{sec:selection_strategy}.
%With this approach, the computational cost is significantly reduced.  
\begin{figure}[t]
    \centering
	\includegraphics[width=\linewidth]{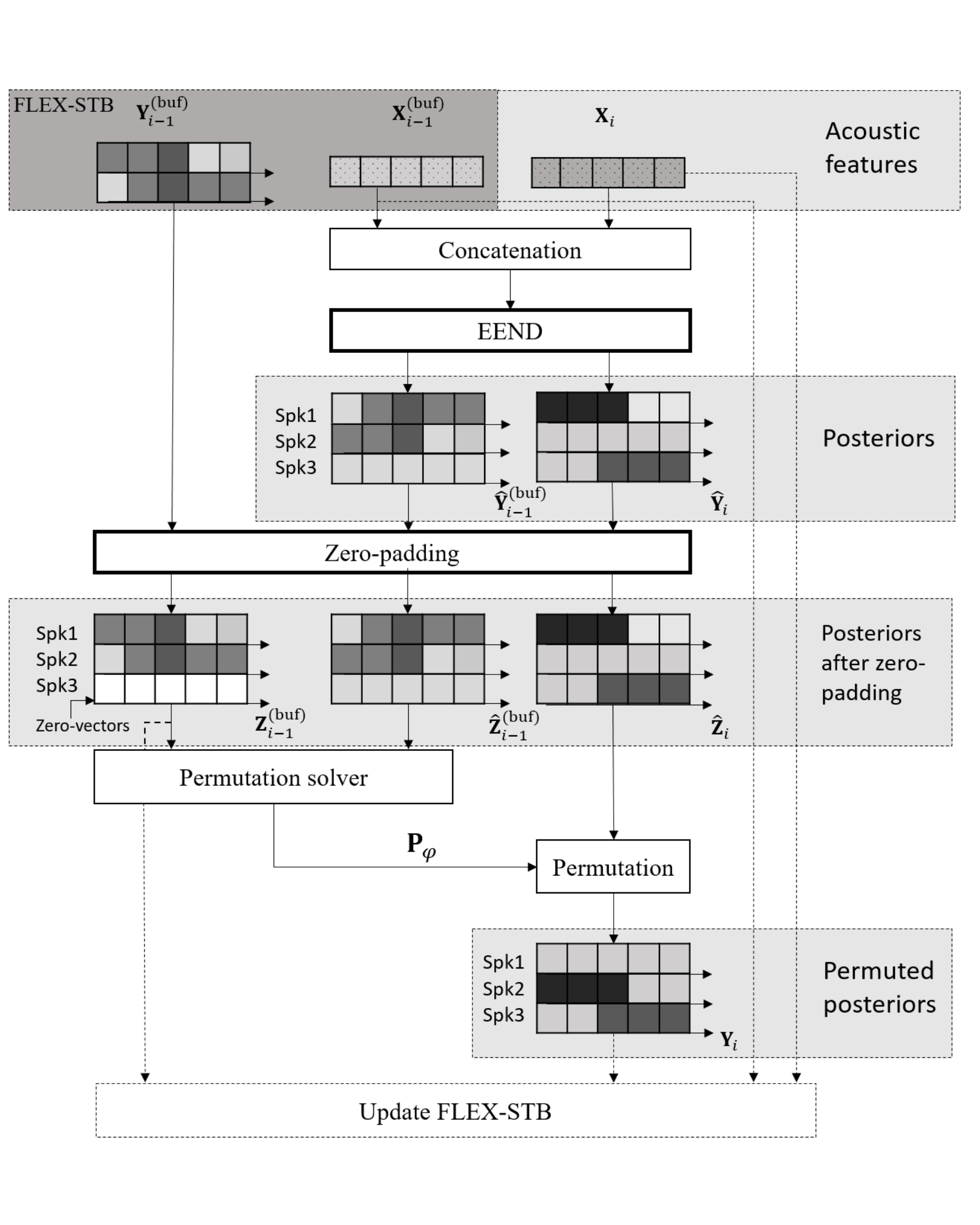}
 	\caption{Proposed speaker-tracing buffer for unknown numbers of speakers. Zero-padding is applied to mitigate the different number of speakers between $\mathbf{Y}^{\text{(buf)}}_{i-1}$ and $\mathbf{\hat{Y}}^{\text{(buf)}}_{i-1}$.}
 	\label{fig:speak_tb}
\end{figure}

\subsection{Speaker-tracing buffer for flexible numbers of speakers (FLEX-STB)}
\label{sec:re-stb}
In this section, we assume that EEND estimates not only speech activities but also the number of speakers $S$, i.e., $\mathrm{EEND}:\mathbb{R}^{D\times T}\rightarrow(0,1)^{S\times T}$.
Firstly, to alleviate the different number of speakers between the buffer $\mathbf{Y}_{i-1}^{\text{(buf)}}$ and the current chunk's output $\mathbf{\hat{Y}}_{i}$, the posterior of the no-speech-activity speaker is considered as zero so that the zero-padding function is applied as follows:
\begin{align}
    \mathbf{Z}^\text{(buf)}_{i-1} &= \mathsf{ZeroPadding}\left(\mathbf{Y}^\text{(buf)}_{i-1}, S_i\right), \label{eq:zero1}\\
    \left[\mathbf{\hat{Z}}_{i-1}^{\text{(buf)}}, \mathbf{\hat{Z}}_i\right] &= \mathsf{ZeroPadding}\left(\left[\mathbf{\hat{Y}}^\text{(buf)}_{i-1}, \hat{\mathbf{Y}}_i\right], S_i\right), \label{eq:zero2}
\end{align}
where $S_i = \max(S_{i-1}, S_i)$ and $\mathsf{ZeroPadding}(\mathbf{A},S)$ appends row zero vectors to $\mathbf{A}$ so that the first dimension becomes $S$.
Next, the speaker permutation $\mathbf{P}_\psi$ for the current chunk is calculated between $\mathbf{Z}_{i-1}^\text{(buf)}$ and $\mathbf{\hat{Z}}^{\text{(buf)}}_{i-1}$ using \autoref{eq:corr}.
%\begin{equation}
%    \psi = \argmax_{\phi \in \mathrm{Perm}(S_t)} %\mathrm{Corr}(\mathbf{Z}_{i-1}^\text{(buf)}, \mathbf{P}_\phi %\mathbf{\hat{Z}}^{\text{buf}}_{i-1}),
%\end{equation}
Then, the output for the current chunk is permuted as follows:
\begin{equation}
    \mathbf{Y}_i = \mathbf{P}_\psi \mathbf{\hat{Z}}_i,
\end{equation}
where $\mathbf{Y}_i$ is the final diarization result of the chunk $i$.
After that, at most $L_\text{max}$ time indexes $\mathcal{T}\subseteq\left\{1,\dots,L_{i-1}+\Delta\right\}$ are selected based on the concatenated outputs $\left[\mathbf{Z}^\text{(buf)}_{i-1}, \mathbf{Y}_i\right] \in (0,1)^{S_i \times (L_{i-1} + T)}$, and the FLEX-STB is updated as follows:
\begin{align}
    \mathbf{X}^\text{(buf)}_i = \left[ \mathbf{x}_\tau \mid \tau \in \mathcal{T} \right], \;
    \mathbf{Y}^\text{(buf)}_i = \left[ \mathbf{y}_\tau \mid \tau \in \mathcal{T} \right],
\end{align}
where $\mathbf{x}_\tau$ is the $\tau$-th column vector of $[\mathbf{X}^\text{(buf)}_{i-1}, \mathbf{X}_i]$, $\mathbf{y}_\tau$ is the $\tau$-th column vector of $[\mathbf{Z}^\text{(buf)}_{i-1}, \mathbf{Y}_i]$.
The frame selection strategies are described in \autoref{sec:selection_strategy}.

\subsection{Selection strategy}
\label{sec:selection_strategy}
When the number of accumulated features becomes larger than the buffer size $L_\text{max}$, a selection strategy is needed to keep relevant features that contain the speaker permutation information from $\left[\mathbf{X}^\text{(buf)}_{i-1}, \mathbf{X}_i\right]$ and $\left[\mathbf{Z}^\text{(buf)}_{i-1}, \mathbf{Y}_i\right]$.
In this section, four selection functions are proposed for flexible numbers of speakers. 

\begin{itemize}
    \item \textbf{Uniform sampling}: 
        Uniform distribution sampling is applied to extract $L_\text{max}$ frames. 
    \item \textbf{First-in-first-out (FIFO)}: The most recent $L_\text{max}$ features and the corresponding diarization results are stored in the buffer, which follows the first-in-first-out manner.
    \item \textbf{Kullback-Leibler divergence based selection}:
        We utilize the Kullback-Leibler divergence (KLD) to measure the difference between two probability distributions: the speaker activities distribution and the uniform distribution at time $t$, which can be represented as follows: 

        \begin{align}
            \text{KLD}_t&= \sum_{s=1}^{S_i}p_{s,t}\log{\frac{p_{s,t}}{q_{s,t}}},\\
        p_{s,t}&=\frac{r_{s,t}}{\sum_{s'=1}^{S_i}r_{s',t}},\\
        q_{s,t}&=\frac{1}{S_i},
        \end{align}
        where $\left[\mathbf{Z}^\text{(buf)}_{i-1}, \mathbf{Y}_i\right] = (r_{s,t})_{\substack{1\leq t\leq(L_{i-1}+\Delta)\\1\leq s\leq{S_i}}}$ is the posteriors from EEND with FLEX-STB and $q_{s,t}$ is the uniform distribution.
        %Small KLD value refers to the consistent distribution tendency to the uniform distribution which tends to indicate that only one speaker speaker exists at time $t$. 
        Top $L_\text{max}$ samples with the highest KLD values are selected from $\left[\mathbf{Z}^\text{(buf)}_{i-1}, \mathbf{Y}_i\right]$ and the corresponding $\left[\mathbf{X}^\text{(buf)}_{i-1}, \mathbf{X}_i\right]$.
    \item \textbf{Weighted sampling using KLD}: The combination of uniform sampling and KLD based selection. $L_\text{max}$ features are randomly selected with the probabilities which are proportional to $\text{KLD}_t$.
\end{itemize}

\begin{table*}[t]
    \centering
    \caption{DERs (\%) of online EEND-EDA with chunk size $\Delta=\SI{1}{\second}$ using FLEX-STB and offline EEND-EDA with chunk size $\Delta=\infty$. 
    %We varied buffer size $L_\text{max}$ from $\SI{10}{\second}$ to $\SI{100}{\second}$, but fixed the chunk size with four buffer selection strategies for online EEND-EDA. 
    \textit{Note that all results are based on the estimated number of speakers, including the overlapping regions without oracle SAD}.  }
    \label{tab:res_buf}
        \resizebox{\linewidth}{!}{%
        \begin{tabular}{@{}lcccccccc@{}}
            \toprule
           &\multicolumn{6}{c}{\bf{Online} (\bf{$\Delta=\SI{1}{\second}$})}&\\
           \cmidrule(lr){2-7} 
           &\multicolumn{3}{c}{CALLHOME} &\multicolumn{3}{c}{DIHARD II}
           &\multicolumn{2}{c}{\bf Offline ($\Delta=\infty$)}\\
           \cmidrule(lr){2-4}\cmidrule(lr){5-7}\cmidrule(l){8-9}
          &$L_\text{max}=\SI{10}{\second}$
           &$L_\text{max}=\SI{50}{\second}$
           &$L_\text{max}=\SI{100}{\second}$ &$L_\text{max}=\SI{10}{\second}$
           &$L_\text{max}=\SI{50}{\second}$
           &$L_\text{max}=\SI{100}{\second}$ &\multicolumn{1}{c}{CALLHOME}&\multicolumn{1}{c}{DIHARD II}\\
       \cmidrule(r){1-7}\cmidrule(l){8-9}
           %\cmidrule(l){1-1}\cmidrule(lr){2-2}\cmidrule(l){3-3}\cmidrule(lr){4-4}\cmidrule(l){5-5}\cmidrule(lr){6-6}\cmidrule(l){7-7}\cmidrule(l){8-8}\cmidrule(l){9-9}
            FLEX-STB with selection strategy \\
            \hspace{3mm}Uniform sampling            & 27.6 & 20.2 & 19.3 & 52.4  & 39.3 & 36.8 & - & - \\
             \hspace{3mm}FIFO                        & 29.5 & \bf19.4 & \bf19.1 & 57.2  & 41.1  & 37.0 & - & - \\
            \hspace{3mm}KLD selection               & 30.0 & 22.3 & 20.9 & 52.6 & 40.8  & 37.7 & - & - \\
            \hspace{3mm}Weighted sampling using KLD & \bf26.6 & 20.0 & 19.5 & \bf50.3  & \bf37.9  & \bf36.0 & - & -\\ 
           \cmidrule(r){1-7}\cmidrule(l){8-9}
            %The best result in (1)-(4) & \bf26.6 & \bf50.3 & \bf19.4 & \bf37.9 & $\bf19.1$ & $\bf36.0$ & $\bf15.3$ & $\bf32.9$ \\ 
            Without FLEX-STB& - & - & - & - & - & - & \bf15.3 & \bf32.9 \\
            \bottomrule
            \end{tabular}%
            }
\end{table*}

\section{Experiment}
\label{sec:exp}

\subsection{Data}
We generated 100k simulated mixtures of one to four speakers following the procedure in \cite{horiguchi2020end} using Switchboard-2 (Phase I, II, III), Switchboard Cellular (Part 1, 1), and the NIST Speaker Recognition Evaluation datasets (SRE).
Additionally, we added noises from the MUSAN corpus \cite{snyder2015musan} and room impulse responses (RIRs) from the Simulated Room Impulse Response Database \cite{ko2017study}.
These simulated mixtures were used for training the EEND-based model. 
Two real conversation datasets: the CALLHOME \cite{callhome} and the DIHARD II \cite{ryant2019second} were prepared for evaluation. 

\subsection{Experiment setting}
In this paper, we evaluated the proposed method on the offline EEND-EDA model. 
The EEND-EDA model was trained with four Transformer encoder blocks and 256 attention units containing four heads \cite{horiguchi2020end}.
We firstly trained the model using a two-speaker dataset for 100 epochs and then finetuned with the concatenation of one- to four-speaker simulated datasets for 25 epochs.
Finally, EEND-EDA model was finetuned using a development set of CALLHOME, or DIHARD II, respectively.

We evaluated all systems with the diarization error rate (DER) metric in both overlapping and non-speech regions.
A collar tolerance of 250 ms was applied at the start and end of each segment for the CALLHOME dataset.
Following the regulation of the second DIHARD challenge \cite{ryant2019second}, we did not use collar tolerance for the DIHARD II dataset.

\subsection{Results}
\subsubsection{Effect of selection strategies and buffer size}
\autoref{tab:res_buf} shows the effect of the selection strategies and the buffer size of the FLEX-STB on the EEND-EDA model in the left part.
Experiment conditions varied from four selection methods with buffer sizes equal to $\SI{10}{\second}$, $\SI{50}{\second}$ and $\SI{100}{\second}$ but fixed the chunk size $\Delta$ to \SI{1}{\second}.
All results were calculated with the estimated number of speakers including the \textit{overlapping regions without oracle sound activity detection (SAD)}. 
It is shown that incremental buffer size which provides more input information improved the accuracy regardless of the selection strategies.
Regarding the selection strategies, on most cases weighted sampling using KLD outperformed other strategies on both datasets.
The best results from online system are \SI{19.1}{\percent} and \SI{36.0}{\percent} for CALLHOME and DIHARD II, respectively. 

\subsubsection{Comparison with the offline EEND-EDA system}
We also compared the performance of our proposed online and baseline offline systems in \autoref{tab:res_buf}.
The input of the offline EEND-EDA system is the whole recording during inference while that for the online system is the \SI{1}{\second} chunk.
Compared with the offline system, DERs of the online system increases by \SI{3.8}{\percent} and \SI{3.1}{\percent} on these two datasets, which would be acceptable degradation by considering the benefit of streaming diarization.
The performance degradation is supposed to come from the mismatch between the offline model which was trained with fixed large chunk size and the online mechanism whose input sizes are incrementally increased. 

\subsubsection{Comparison with other online diarization systems}
%We also compared our proposed method with the recently proposed BW-EDA-EEND\cite{han2020bw}.
First, we compared our method with the recently proposed BW-EDA-EEND \cite{han2020bw} on the CALLHOME dataset.
In order to compare with BW-EDA-EEND in the same condition, we evaluated our method with a \SI{10}{\second} chunk size.
As shown in \autoref{tab:bw-eda}, in a \SI{10}{\second} chunk size and estimated SAD condition, our proposed method outperforms the BW-EDA-EEND on all speaker-number cases on the CALLHOME dataset. 

Next, we compared our proposed method with other systems in more realistic scenario, i.e., DIHARD II.
For a fair comparison with other online methods, we follow the DIHARD II track 1, where \textit{the oracle SAD information is provided}.
We used the oracle SAD information to filter out non-speech frames of the estimated diarization result. 
\autoref{tab:compare} shows the comparison with other systems. 
The proposed online EEND-EDA with FLEX-STB achieved a DER of \SI{25.8}{\percent}, which outperformed the UIS-RNN-SML, and is comparable to the \textit{offline} DIHARD II baseline.
%As shown in \autoref{tab:compare}, the proposed online EEND-EDA with FLEX-STB achieved a DER of 25.8\%, which outperformed the  UIS-RNN-SML\cite{fini2020supervised}, even better than the offline baseline \cite{ryant2019second} which uses the agglomerative clustering and PLDA scoring.

\subsubsection{Real-time factor and latency}
%Our experiment was conducted on an Intel\textsuperscript{\textregistered} Xeon\textsuperscript{\textregistered} CPU E52697A v2 @ 2.60GHz using one thread. 
Our experiment was conducted on one NVIDIA Tesla P100 GPU.
To calculate the average computing time of one buffer, we filled the buffer with dummy values for the first iteration to keep the buffer size always the same among chunks.
The real-time factor was equal to 0.13 when we applied FLEX-STB to EEND-EDA with chunk size equal to $\SI{1}{\second}$, and a buffer size of $\SI{100}{\second}$.
This means that the average computation duration of a \SI{1}{\second} chunk was \SI{0.13}{\second} which is acceptable for the online processing. 

\begin{table}[t]
    \caption{DERs (\%) of each number of speakers on the CALLHOME dataset with \SI{10}{\second} chunk size. \textit{Both estimated the number of speakers and included the overlapping regions without using oracle SAD.}}
    \centering
    \begin{tabular}{@{}lccc@{}}
    \toprule
     &\multicolumn{3}{c}{Number of speakers} \\ \cmidrule(l){2-4}
    Method     &   2 & 3 & 4 \\ \midrule
    %\hspace{2mm} 
    BW-EDA-EEND \cite{han2020bw} & 11.8 & 18.3 & 26.0 \\ 
    %\hspace{2mm} 
    EEND-EDA w/ FLEX-STB  & \bf 10.0 & \bf14.0  & \bf 21.1  \\
    \bottomrule
    \end{tabular}
    \label{tab:bw-eda}
\end{table}

\begin{table}[]
    \caption{DERs (\%) on DIHARD II dataset \textit{computed by using oracle SAD including overlapping regions}. Online systems with STB were evaluated in a \SI{1}{\second} chunk size $\Delta$ and \SI{100}{\second} buffer size $L_{\mathrm{max}}$.}
    \centering
    \begin{tabular}{@{}lc@{}}
    \toprule
    Method & DER \\ \midrule
      DIHARD-2 baseline (offline) \cite{ryant2019second} & 26.0 \\
      %UIS-RNN & 30.9 \\
      UIS-RNN-SML \cite{fini2020supervised}   &  27.3\\
      EEND-EDA w/ FLEX-STB    & \bf25.8 \\
    \bottomrule
    \end{tabular}
    \label{tab:compare}
\end{table}

\section{Conclusion}
In this paper, we proposed an online streaming speaker diarization method that handles overlapping speech and flexible numbers of speakers.
A speaker tracing buffer for flexible numbers of speakers was proposed to mitigate the different number of speakers among chunks.
Experimental results showed that the proposed online system achieves comparable results with the offline method and better results than the BW-EDA-EEND online method.
%On the DIHARD II track 1 task, it achieved better results than the offline baseline and also the online UIS-RNN method. 
One of our future studies is to incorporate various extensions developed at the recent DIHARD III challenge, including semi-supervised training and model fusion \cite{horiguchi2021hitachi}.

\bibliographystyle{IEEEtran}

\bibliography{main}

% \begin{thebibliography}{9}
% \bibitem[1]{Davis80-COP}
%   S.\ B.\ Davis and P.\ Mermelstein,
%   ``Comparison of parametric representation for monosyllabic word recognition in continuously spoken sentences,''
%   \textit{IEEE Transactions on Acoustics, Speech and Signal Processing}, vol.~28, no.~4, pp.~357--366, 1980.
% \bibitem[2]{Rabiner89-ATO}
%   L.\ R.\ Rabiner,
%   ``A tutorial on hidden Markov models and selected applications in speech recognition,''
%   \textit{Proceedings of the IEEE}, vol.~77, no.~2, pp.~257-286, 1989.
% \bibitem[3]{Hastie09-TEO}
%   T.\ Hastie, R.\ Tibshirani, and J.\ Friedman,
%   \textit{The Elements of Statistical Learning -- Data Mining, Inference, and Prediction}.
%   New York: Springer, 2009.
% \bibitem[4]{YourName17-XXX}
%   F.\ Lastname1, F.\ Lastname2, and F.\ Lastname3,
%   ``Title of your INTERSPEECH 2021 publication,''
%   in \textit{Interspeech 2021 -- 20\textsuperscript{th} Annual Conference of the International Speech Communication Association, September 15-19, Graz, Austria, Proceedings, Proceedings}, 2020, pp.~100--104.
% \end{thebibliography}

\end{document}